\documentclass[conference]{IEEEtran}
\IEEEoverridecommandlockouts
\usepackage{cite}
\usepackage{amsmath,amssymb,amsfonts}
\usepackage{algorithmic}
\usepackage{graphicx}
\usepackage{textcomp}
\usepackage{xcolor}
\usepackage{color}
\usepackage{fancyhdr}
\usepackage{booktabs}
\usepackage{tabu}
\usepackage{bm}
\usepackage{tabularx, booktabs, array, amssymb, bbding}
\usepackage{caption2}

\def\BibTeX{{\rm B\kern-.05em{\sc i\kern-.025em b}\kern-.08em
    T\kern-.1667em\lower.7ex\hbox{E}\kern-.125emX}}
\begin{document}

\title{Cross-Modal Self-Attention Distillation for Prostate Cancer Segmentation\\
\thanks{\IEEEauthorrefmark{7}: Corresponding author: Ye Luo, Jianwei Lu.}
\thanks{\IEEEauthorrefmark{2}: Guokai Zhang and Xiaoang Shen contributed equally to this work.}
}

\author{\IEEEauthorblockN{Guokai Zhang\IEEEauthorrefmark{1}\IEEEauthorrefmark{2},
Xiaoang Shen\IEEEauthorrefmark{3}\IEEEauthorrefmark{2},
Ye Luo\IEEEauthorrefmark{3}\IEEEauthorrefmark{7}, 
Jihao Luo\IEEEauthorrefmark{4}, 
Zeju Wang\IEEEauthorrefmark{3},\\
\vspace{0.1cm}
Weigang Wang\IEEEauthorrefmark{5},
Binghui Zhao\IEEEauthorrefmark{6},and 
Jianwei Lu\IEEEauthorrefmark{3}\IEEEauthorrefmark{7}}
\IEEEauthorblockA{\IEEEauthorrefmark{1}School of Optical-Electrical and Computer Engineering,University of Shanghai for Science and Technology, Shanghai, China}
\IEEEauthorblockA{\IEEEauthorrefmark{3}School of Software Engineering,Tongji University,Shanghai, China}
\IEEEauthorblockA{\IEEEauthorrefmark{4}School of Computing,National University of Singapore,Singapore}
\IEEEauthorblockA{\IEEEauthorrefmark{5}Department of Radiology,Shanghai Fire Corps Hospital,Shanghai, China}
\IEEEauthorblockA{\IEEEauthorrefmark{6}Department of Radiology,Shanghai Tenth People’s Hospital,Shanghai, China}
\IEEEauthorblockA{Emails:\hspace{0.5em}\{yeluo,jwlu33\}@tongji.edu.cn}}

\maketitle

\thispagestyle{fancy}            
\fancyhead{}                     
\chead{\textmd{2020 IEEE International Conference on Bioinformatics and Biomedicine (BIBM)}}                
\cfoot{\quad}                    
 
\renewcommand{\headrulewidth}{0pt}      
\renewcommand{\footrulewidth}{0pt}
 
\pagestyle{empty} 




\begin{abstract}
Automatic segmentation of the prostate cancer from the multi-modal magnetic resonance images is of critical importance for the initial staging and prognosis of patients. However, how to use the multi-modal image features more efficiently is still a challenging problem in the field of medical image segmentation. In this paper, we develop a cross-modal self-attention distillation network by fully exploiting the encoded information of the intermediate layers from different modalities, and the extracted attention maps of different modalities enable the model to transfer the significant spatial information with more details. Moreover, a novel spatial correlated feature fusion module is further employed for learning more complementary correlation and non-linear information of different modality images. We evaluate our model in five-fold cross-validation on $358$ MRI with biopsy confirmed. Extensive experiment results demonstrate that our proposed network achieves state-of-the-art performance.
\end{abstract}


\section{Introduction}

Prostate cancer (PCa) has been one of the most common cancer and second leading cause death in the world of the men \cite{Siegel}.
Accurate segmentation of PCa from multi-modal magnetic resonance imaging (MRI) plays an essential role in the diagnosis and treatment of PCa disease. 
However, manual segmentation is a time-consuming and error-prone task. 
Thus, developing an automatic and efficient method to segment PCa is in a high demand that can help the radiologist alleviate the workload in clinical diagnosis.

Automatic segmentation of the PCa from the MRI is usually a very challenging task. 
Many methods have been proposed to automatically segment PCa from MRI. The early attempts for PCa detection are mainly focused on the hand-crafted features \cite{Geert,Rania,Khalvati,McGarry}. Although those methods have achieved promising results, they could be limited by the empirical and subjective ways to extract features.  
With the remarkable performance of convolutional neural networks (CNNs), many deep-learning-based methods have gained excellent results in MRI analysis tasks. Considering to utilize rich data for accurate diagnosis, the recent state-of-the-art medical image segmentation methods benefited from multi-modal learning face a common issue: the fusion of features from different modalities. Specifically, early fusion \cite{Kiraly,wangY,caoR}, late fusion \cite{Wu}\cite{Nie} and multi-layer fusion \cite{LiC}\cite{Dolz} are the three mainstream fusion strategies.  
While the early fusion strategy which directly fusing images as the raw input ignores the discrepancies among different modality images, late fusion strategy gets its superiority via high-level feature fusion but pays the cost of severe network complexity increment.
\begin{figure*}
\centerline{\includegraphics[scale=0.5]{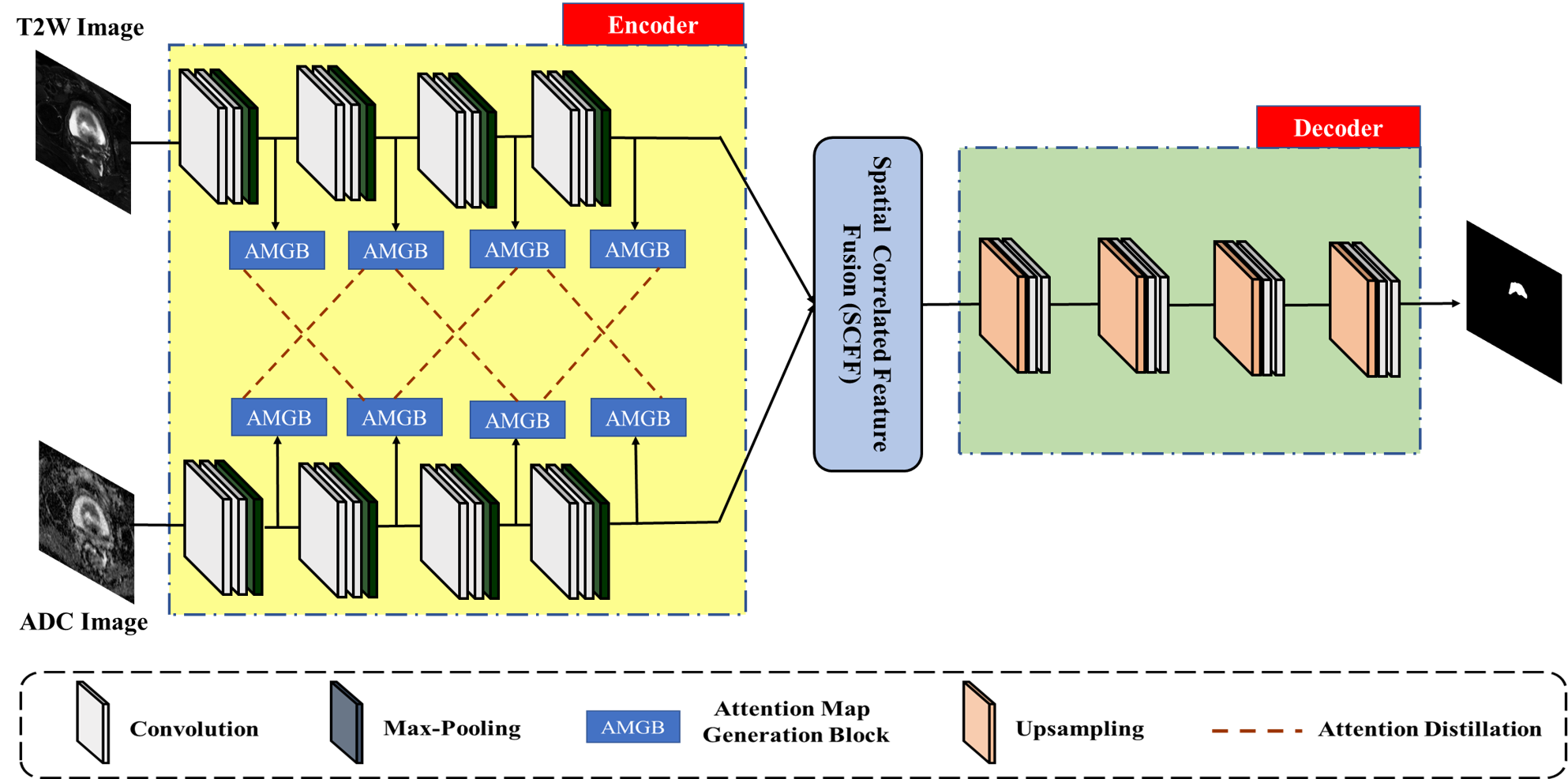}}

\caption{Overview of our CSAD network. Two different modality MRI (i.e. T2W and ADC) are sent into a two-stream Encoder sub-network. During training, the input images and corresponding masks are used to optimize the encoder-decoder network for segmentation task.
At each intermediate layer of the Encoder sub-network, a group of AMGBs are added to implement the cross-modal self attention distillation. The output features of those two modalities from the Encoder are then fused by the SSFF module and the final prediction mask is generated by the Decoder.} 
\label{main_fig}
\end{figure*}

Multi-layer fusion strategy is hence proposed to hierarchically fuse features from each modality layer-wisely \cite{ZhangS,LiC,Dolz,ChenY,Jiaz}. Although the multi-layer fusion strategy has achieved great progresses, most of works treat all modality equally except for \cite{LiC} which employs the knowledge distillation (KD) to make the assistant modality generate supervision information to help the master modality learning. Based on the same motivation to distinguish modalities, an activation-based attention is generated and used to weight features extracted from each modality for brain tumor segmentation \cite{ZhangS}. Instead of leveraging attention to weight different modalities or performing KD on two networks, in this paper, we propose to distill the attention-based knowledge across modalities within one network, by which the significant spatial information of the lesion region can be directly learned layer-wisely. And an attention map used as the soft label to guide the network learning is not new and it gets broad applications to other tasks in existing work \cite{Zagoruyko}\cite{HouY}.

Different from the aforementioned methods and based on the assumption that even at different modality or network layer significant spatial information about the lesion region should be consistent, we propose a cross-modal self-attention distillation (CSAD) network for prostate cancer segmentation from a whole MRI, which transfers attention-based knowledge among different modalities and then features learned from different modalities are fused with the designed spatial correlated feature fusion (SCFF) module.
The major strengths of our CSAD network are:

1) By distilling the attention-based knowledge across modalities, the cross-modal reciprocal correlations are exploited and the useful information from different modalities are implicitly fused into the layer-wisely learned features. 

2) Instead of using the attention map to describe the relationships of various modalities, we learn the significant spatial information of the lesion directly by the attention map and enforce the representation learning by the interlaced CSAD. 

3) Different from the late fusion strategy, a SCFF module is designed and incorporated in the CSAD network, which allows the network to learn more complementary correlation and non-linear information from different modality images.


\section{Related Work}

\subsection{Hand-crafted feature methods for PCa Segmentation}
The early works for PCa segmentation are mainly focused on the hand-crafted feature selection methods which utilizing the predefined image features to construct a feature-empirical model.
For example, Geert et al. \cite{Geert} proposed a PCa detection system, which used the multi-atlas-based and local maxima detection to detect the PCa candidates.
Liu et al. \cite{LiuX} proposed a novel prostate cancer segmentation method based on the fuzzy markov random fields in an unsupervised manner.
To consider the interpretability and performance of the model simultaneously, Wang et al. \cite{WangY} employed a stacking-based ensemble learning approach with the random forest to achieve the prostate cancer detection.
In \cite{McGarry}, it predicted the lumen and epithelium density values of each MRI and then located the high-sensitive PCa region automatically.
Then, Khalvati et al. \cite{Khalvati} designed a MPCaD which leveraged the multi-scale feature extraction ability to achieve the prostate cancer detection and localization.
While those hand-crafted based methods have achieved great success, it could have the limitation of selecting the features empirically and subjectively, and thus hinder the designed model to learn a more comprehensive representation of PCa.

\subsection{Deep learning for PCa Segmentation}
While there are tremendous of deep learning methods on the medical image analysis, the approaches based on CNN for the PCa segmentation from the whole MRI is still lagging.
In \cite{AlkadiR}, Alkadi et al. proposed a PCa detection model by using the T2W images, it segmented the prostate region and PCa tissues by exploiting the 3D spatial information.
However, adopting the single modality of MRI may ignore the reciprocal information of different modalities and then hinder the model to achieve a better segmentation performance.
To address this challenge, in \cite{Kiraly}, Kiraly et al. designed a multi-channel encoder-decoder network from multi-parametric MRI to achieve the detection and classification of PCa, and then the k-fold experiments were conducted on multi-parametric MRI, and the experimental results demonstrated that the designed model could achieve $83.4\%$ AUC score.
Yang et al. \cite{YangX} employed a multi-path network based on multi-parametric MRI with a weakly-supervised learning to detect the aggressive PCa lesions automatically. Extensive experiments on $402$ lesions showed that the proposed method could achieve state-of-the-art at that time.
Kohl et al. \cite{Kohl} first proposed an adversarial network to segment the aggressive PCa region where the segmentation result was refined by discriminating the expert and predicted annotations.
Though those methods have gained satisfied results, the way to distill the attention-based knowledge across different modalities is still not attempted, and how to full utilization cross different modalities is still a challenge.


\section{Methodology}
The overview of our CSAD network is shown in Fig. \ref{main_fig}. 
In the Encoder, features from two modalities: T2-weighted (i.e. T2W) and Apparent diffusion coefficient (i.e. ADC) are extracted via a two-stream sub-network. Within each branch of the sub-network, a group of attention map generation blocks (i.e. AMGBs) are placed onto the intermediate layers after every encoding stage. Attention distillation is performed onto two AMGBs from T2W and ADC in an interlaced way. To fuse the output features from two modalities, the SCFF module is adopted. And the final prediction mask is generated by the Decoder. 

\subsection{Cross-Modal Self-Attention Distillation}
\subsubsection{Attention Map Generation Block}
In order to generate attention maps from the features of different layers, as shown in Fig. \ref{main_fig}, an AMGB is built and integrated into every intermediate layer of the two-stream Encoder sub-network. 
Generally, there are two attention map generation approaches: attention-based and gradient-based \cite{Zagoruyko}, and we employ the activation-based attention in our proposed CSAD network as \cite{HouY} did. To calculate the activation-based attention map, some notations are introduced first. Denote $F_m^k \in R^{C_m^k \times W_m^k \times H_m^k }$ the feature maps output from the $m$-th encoder block in $k$-th modality stream, here $m\in [1, 5]$, $k \in [1,2]$, and $C_m^k$, $W_m^k$ and $H_m^k$ are the channel number, the height and the width of $F_m^k$, respectively. Then denote $S_m^k \in R^{W_m^k \times H_m^k }$ and  $A_m^k \in R^{W_m^k \times H_m^k }$ as the generated spatial map and the attention map for $F_m^k$, respectively. We have $A_m^k = \mathcal{G}(S_m^k)$, where $\mathcal{G}(\cdot)$ is the spatial softmax operation. To generate the spatial map is equivalent to finding an activation-based mapping function $\mathcal{R}:F_m^k \rightarrow S_m^k$. Various mapping functions are proposed, and one of them reveals that the absolute value of the element in $F_m^k$ is considered as the importance of this element in the final output \cite{Zagoruyko}. And this idea got comprehensive comparisons in the later researches and it is further confirmed that the spatial map generated by $S_m^k=\sum_{t=1}^{C_m^k}|F_{m,t}^k|^2$ can put more weight to the most discriminative regions. Thus, we calculate our own attention map $A_m^k$ for $F_m^k$ as: 
\begin{equation}
A_m^k=\mathcal{G}(\sum_{t=1}^{C_m^k}|F_{m,t}^k|^2).
\end{equation}
Moreover, considering that the attention maps generated at different layers have different sizes, we upsample the feature maps by $\mathcal{P}(\cdot)$ to the target size before the spatial mapping operation $\mathcal{R}(\cdot)$. Hence, the whole procedure of AMGB $\chi(\cdot)$ can be formulated by the equation as:
\begin{equation}
\chi(\cdot)=\mathcal{G}(\mathcal{R}(\mathcal{P}(\cdot))).
\end{equation}

\subsubsection{Attention Distillation}
\label{attention_dis}
The concept of self-attention distillation is originated from \cite{HouY}, in which attention maps extracted from various layers of a model encode rich contextual information that can be used as a form of free supervision for further representation learning through performing layer-wise attention distillation within the network itself. Therefore, in our CSAD network, after obtaining the attention maps via each AMGB at different layer, the CSAD is performed onto the consecutive layers such that the significant spatial information can transfer not only within individual modality but also across modalities. Specifically, significant spatial information can be distilled from an attention map of one modality (i.e. T2W) and then be used to facilitate attention learning of the other modality (i.e. ADC). As shown in Fig.~\ref{main_fig},  the dashed line connects an AMGB at T2W stream to another AMGB at ADC stream, and vice versa. Moreover, we perform the CSAD in an interlaced way such that the long dependencies of features (e.g. the details at deep layer and the structural information at shallow layer) can be exploited across modalities. Significantly, although there are different connection schemes of the attention distillation paths (e.g. $AMGB_m \rightarrow AMGB_{m+1}$, $AMGB_m \rightarrow AMGB_{m+2}$,etc.), we choose the current connection strategy to balance the performance and the model complexity.  After attention distillation, the two streams of the Encoder should have the agreement onto the knowledge about the lesion, thus the Kullback-Leibler (KL) divergence between the two attention maps from two modalities should be minimized. The attention distillation loss $\mathcal{L}_{CSAD}$ between two successive attention maps of different modalities can be formulated as follows:
\begin{equation}
\label{eq:csad}
\begin{split}
\mathcal{L}_{CSAD}=\sum_{m=1}^{N-1}(\mathcal{L}_{AD}(\chi(F_m^k),\chi(F_{m+1}^{\hat{k}}))+\\
\mathcal{L}_{AD}(\chi(F_{m}^{\hat{k}}), \chi(F_{m+1}^k))),
\end{split}
\end{equation}
where $\mathcal{L}_{AD}$ is a new symmetric version of KL divergence to enhance the alignment of the distilled attention. $N$ is the total number of attention maps in single modality. $k$ and $\hat{k}$ are the indexes of two modality streams. Formally, our attention distillation based loss term $\mathcal{L}_{AD}$ can be formulated as follows:
\begin{equation}\label{loss}
\mathcal{L}_{AD}=\frac{1}{2W H}\sum_{i=1}^{W}\sum_{j=1}^{H}(\mathcal{L}_{KL}(a_{ij}|b_{ij})+\mathcal{L}_{KL}(b_{ij}|a_{ij})),
\end{equation}
where $W$ and $H$ are the width and the height of an attention map, respectively. And $a_{ij}$ and $b_{ij}$ denote the pixel values at location $(i,j)$ of the attention maps $a$ and $b$, respectively. The $\mathcal{L}_{KL}$ in Eq. \ref{loss} is the standard KL divergence.

\subsection{Spatial Correlated Feature Fusion}
The feature fusion strategy is especially crucial to handle the data with various modalities in that we need to preserve the information from individual modality and meanwhile make them complementary to each other. However, directly applying naive feature fusion operation (eg. concatenation, element-wise addition) to features from discrepant modalities may yield defective features and thus incur a degree of performance loss, since the semantic gap between different modality features is often neglected. Hence, even after implicitly fusing features across different modalities via the CSAD network, we investigate the feature relationship between different modalities through a SCFF module as shown in Fig.\ref{main_fig}. 

Let $F_{T_2} \in \mathbb{R}^{C \times H \times W}$ and $F_{ADC} \in \mathbb{R}^{C \times H \times W}$ denote the output features from the last Encoder layer of T2W and ADC stream respectively. We first pass $F_{T_2}$ and $F_{ADC}$ through a same $1 \times 1 $ convolution layer to extract general features $\tilde{F}_{T_2}\in \mathbb{R}^{\tilde{C} \times \tilde{H} \times \tilde{W}}$ and $\tilde{F}_{ADC}\in \mathbb{R}^{\tilde{C} \times \tilde{H} \times \tilde{W}}$ . These general features incorporate aligned spatial position information and context details with the semantic gap eliminated. To fully explore the spatial correlation between $\tilde{F}_{T_2}$ and $\tilde{F}_{ADC}$, we define the spatial correlation matrix $
Z\in\mathbb{R}^{\tilde{H}\tilde{W} \times \tilde{H}\tilde{W}}$ to quantify this relationship:
\begin{equation}
z_{i,j}=\frac{1}{\mathcal{T}}f(x_i, y_j),
\end{equation}
where $x_i \in \mathbb{R}^{\tilde{C}}$ and $y_j \in \mathbb{R}^{\tilde{C}}$ denote the features at the $i$-th and $j$-th position of $\tilde{F}_{T_2}$ and $\tilde{F}_{ADC}$ respectively, and $z_{i,j}$ is the resulting scalar at $(i,j)$ in $Z$.  $\mathcal{T}$ is the normalization factor. The pairwise function $f$ computes a scalar which represents the spatial correlation between $x_i$ and $y_j$. Here, for simplicity, we choose the dot-product similarity as $f$, which is formulated as:
\begin{equation}
f(x_{i},y_{j})=x_{i}^Ty_{j}.
\end{equation}
In this case, we set $\mathcal{T}=\tilde{C}$ to normalize the output scalar. After obtaining the spatial correlation matrix $Z$, a spatial softmax is performed to get a re-weighted spatial correlation map $Z_{\mathcal{G}}$. Through $Z_{\mathcal{G}}$, we can project $F_{T_2}$ and $F_{ADC}$ to a same latent feature space with the spatial correlation between them fully explored. The final fused feature maps $F_{corr}$ can be defined as:
\begin{equation}
F_{corr}=\Omega({\hat{F}_{T_2}}\otimes Z_{\mathcal{G}})\parallel\Omega({\hat{F}_{ADC}}\otimes Z_{\mathcal{G}}^T),
\end{equation}
where $\hat{F}_{T_2} \in \mathbb{R}^{\tilde{C} \times \tilde{H}\tilde{W}}$ and $\hat{F}_{ADC} \in \mathbb{R}^{\tilde{C} \times \tilde{H}\tilde{W}}$ are the reshaped form of $F_{T_2}$ and $F_{ADC}$ to fit the matrix multiplication, the operation of $\otimes$ represents the matrix multiplication. $\Omega(\cdot)$ reshapes the input features from $\mathbb{R}^{\tilde{C} \times \tilde{H}\tilde{W}}$ to $\mathbb{R}^{\tilde{C} \times \tilde{H} \times \tilde{W}}$, and $\parallel$ denotes concatenation operation.
After obtaining the fused feature $F_{corr}$ through SCFF, we deliver it to the Decoder stream for the segmentation prediction.

\subsection{Multi-Modal Learning}
Our compact cross-modal self-attention distillation network is trained with the following hybrid loss function:
\begin{align}
\mathcal{L}_{total} = \mathcal{L}_{Seg} + \alpha\mathcal{L}_{CSAD}.
\end{align}
Here $\alpha$ is the relax parameter adopted to trade off between two sub loss functions (i.e. $\mathcal{L}_{Seg}$ and $\mathcal{L}_{CSAD}$) in $\mathcal{L}_{total}$ respectively.
Note that the $\mathcal{L}_{Seg}$ is the loss function for the segmentation task, which is the combination of the Dice loss $\mathcal{L}_{Dice}$ and the weighted binary cross-entropy loss $\mathcal{L}_{WBCE}$, which can be formulated as:
\begin{equation}
    \mathcal{L}_{Seg} = \mathcal{L}_{Dice}+\beta\mathcal{L}_{WBCE},
\end{equation}
where $\beta$ is the hyper-parameter to balance those two losses.
Instead of using the common binary cross-entropy loss in $\mathcal{L}_{Seg}$, we proceed to the weighted binary cross-entropy loss in that extreme class-imbalance problem extensively exists in our dataset. Thus, the detailed formulations of $\mathcal{L}_{Dice}$ and $\mathcal{L}_{WBCE}$ are:
\begin{equation} \label{dice_loss}
    \mathcal{L}_{Dice} = 1 - \frac{2|X \cap Y| + \sigma}{|X|+|Y|+\sigma}.
\end{equation}
\begin{equation} \label{wbce_loss}
    \mathcal{L}_{WBCE} = \frac{1}{|X|}\sum_{i=1}^{N}[-\lambda p_{i} log\hat{p}_{i} - (1-\lambda)(1-p_{i}) log (1-\hat{p}_i)],
\end{equation} 
where $X$ and $Y$ are the predicted mask and the ground truth mask, respectively. $\sigma$ is a smoothing factor to control the numerical stability, $p_{i}\in Y$ is the label of the $i$-th pixel in the ground truth mask while $\hat{p}_{i} \in X$ is corresponding predicted value, and $\lambda$ is a weight and empirically set to $0.95$ in the following experiments.

\section{Experiment}
\begin{table*}[h]
\setlength{\belowcaptionskip}{5pt}
\fontsize{8}{12}\selectfont
    \caption{The results of ablation studies.}
        \label{tab1}
    \centering
  \begin{tabular}{ p{2.5cm}<{\centering}|m{2cm}<{\centering}|m{2cm}<{\centering}|m{2cm}<{\centering} |m{2cm}<{\centering}|m{2cm}<{\centering}}
\hline
Method & DICE (\%) & Sensitivity (\%) & Precision (\%) & VOE (\%) & RVD (\%) \\
\hline
\hline
Baseline  & 58.2 $\pm$ 1.0 & 67.4 $\pm$ 2.1 & 58.3 $\pm$ 0.9 & 52.2 $\pm$ 3.3 & 29.3 $\pm$ 1.5\\
Baseline+SCFF  & 60.5 $\pm$ 0.7 & 70.3 $\pm$ 2.7 & 58.8 $\pm$ 1.4 & 49.2 $\pm$ 3.4 & 30.8 $\pm$ 2.1\\
Baseline+CSAD  & 62.6 $\pm$ 0.9 & 73.0 $\pm$ 1.9 & 60.4 $\pm$ 0.5 & 46.8 $\pm$ 2.7 & 24.7 $\pm$ 2.8\\
\textbf{Proposed} & \textcolor{black}{\textbf{65.7 $\bm{\pm}$ 1.2}} & \textcolor{black}{\textbf{77.8 $\bm{\pm}$ 2.5}} & \textcolor{black}{\textbf{60.7 $\bm{\pm}$ 0.4}} & \textcolor{black}{\textbf{44.5 $\pm$ 4.0}} & \textcolor{black}{\textbf{17.5 $\pm$ 2.3}} \\
\hline
\end{tabular}
\end{table*}

\subsection{Dataset and Implementation Details}
The experiments are conducted on $358$ MRI provided by a local hospital. The voxel size of the image is $0.9 \times 0.6 \times 3.5 mm^{3}$,  and the whole dataset was collected using the 3.0 Tesla (T) whole-body unit MRI system.  The experimental modalities are T2W and ADC modality images and  the non-grid registration is utilized among different modalities to make images consistent in positions.  
For the two modalities, 2D slices are used as inputs of the model, and all the annotations of the prostate cancer are labelled by two skilled radiologists and then cross-validated by another two radiologists. The result is further confirmed by the biopsy pathology.
The images and masks in our dataset are resized to $256 \times 256$ for saving memory usage. 
To alleviate the over-fitting problem caused by the small dataset size, we extend the existing dataset by adopting several data augmentation techniques including random rotation, zoom, horizontal/vertical shift and Gaussian Blur. Our model is implemented with PyTorch. We use SGD optimizer with a fixed batch size of 2 to train the network on a NVIDIA GTX 1080Ti GPU for 100 epochs. The hyper-parameters $\alpha$, $\beta$ and $\sigma$ are empirically set to $1$, $0.1$ and $1$ respectively. All the experiments follow a standard five-fold cross-validation procedure.

\subsection{Evaluation Metrics}
In the experiment, we have adopted five commonly accepted metrics to evaluate image segmentation results, including Dice Similarity Coefficient (Dice), Sensitivity, Precision, Volumetric Overlap Error (VOE) and Relative Volume Difference (RVD). We do not embrace accuracy as one of our evaluation metrics because of the extreme foreground-background (i.e. cancer-noncancer) class imbalance in the MRI cancer image.

Dice score is a measure function which is usually used to calculate the similarity or overlap of two samples. Its definition is shown as below:

\begin{equation}
    Dice = 1 - \frac{2|A \cap B|}{|A|+|B|}.
\end{equation}

and VOE is defined as follows:
\begin{equation}
    VOE = 1 - \frac{|A \cap B|}{|A \cup B|} \times 100(\% ).
\end{equation}

RVD is used to describe the relative difference of two object volumes which can be defined as follows:
\begin{equation}
    RVD = \frac{|B|-|A|}{|A|} \times 100(\% ).
\end{equation}
Here A and B are the predicted mask and the ground truth, respectively. $|.|$ means to count the number of pixels.  Specifically, the lower score of VOE and RVD gains, the better segmentation performance would be.
Other metrics (e.g. Precision and Sensitivity) are defined as follows:
\begin{equation}
Precision = \frac{TP+TN}{TP + FN + TN+FP}.
\end{equation}
\begin{equation}
Sensitivity = \frac{TP}{TP + FN}.
\end{equation}
Here TP refers to the number of the true-positives samples; TN represents the number of true-negative samples; FP is the number of the false positives and FN denotes to the number of the false negatives.

\subsection{Ablation Study}
We conduct the following experiments to validate the effectiveness of each component in our proposed method. The \textbf{Baseline} is pure the two-stream Encoder sub-network and the Decoder module without the CSAD and the SCFF modules but with normal concatenation for feature fusion. `\textbf{Baseline+CSAD}' means to add the CSAD mechanism without the SCFF module to the basline, while `\textbf{Baseline+SCFF}' retains the SCFF module but without CSAD. And `\textbf{Proposed}' incorporates our complete design. The detailed comparison results can be referred to Table~\ref{tab1}.  From this table, significant performance improvement can be seen by adding CSAD to the baseline. While the improvement is not so obvious by adding SCFF only compared to `\textbf{Baseline+CSAD}'. And our method incorporated with both CSAD and SCFF achieves the best performance.
\begin{figure*}[t]
\centerline{\includegraphics[scale=0.5]{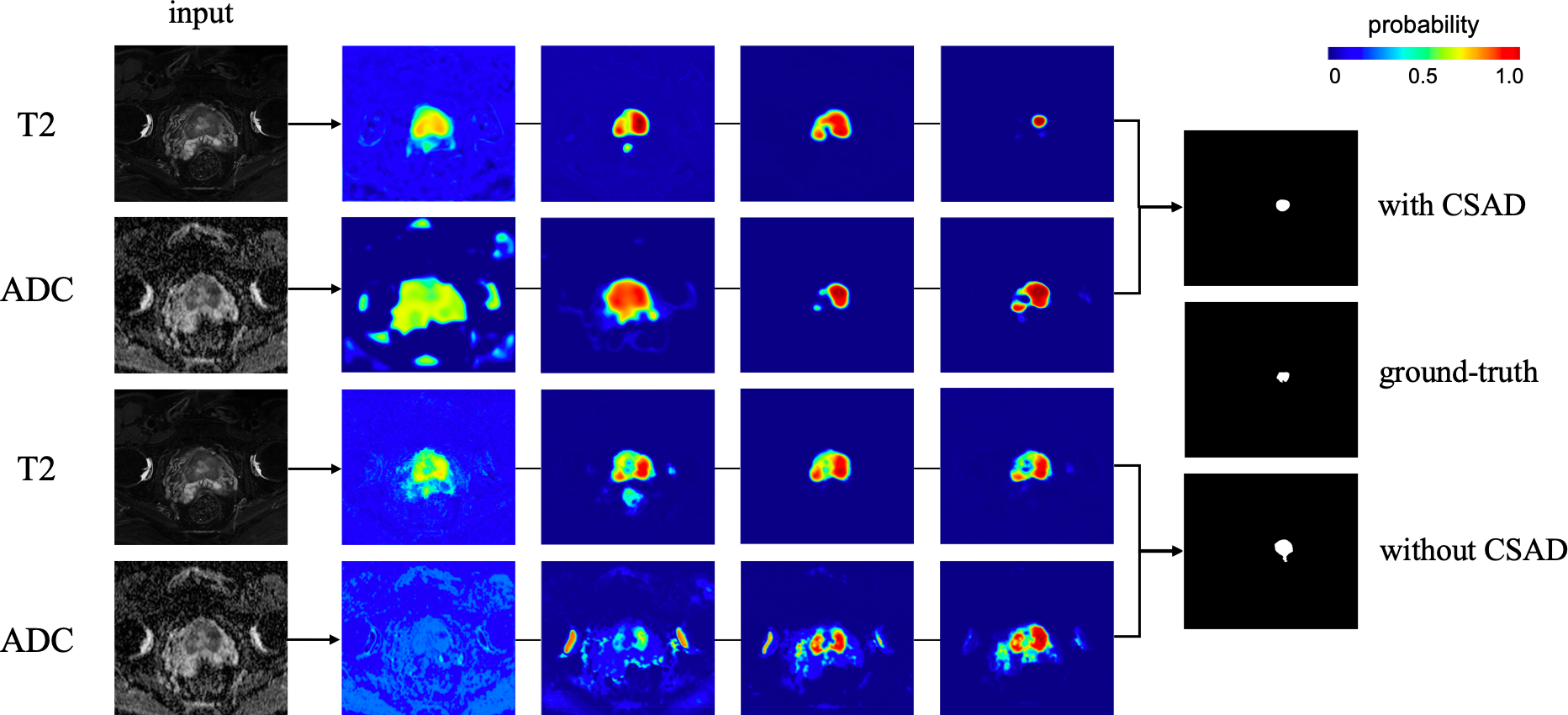}}
\caption{A visual comparison of the attention maps generated with (top two rows) and without (bottom two rows) the CSAD. The input images of the same modality are identical.} \label{fig_am}
\end{figure*}
\doublerulesep=0.4pt
\begin{table*}[h]
\setlength{\belowcaptionskip}{5pt}
\fontsize{8}{12}\selectfont
    \caption{Effectiveness of different connection schemes.}
        \label{tb_connection}
    \centering
  \begin{tabular}{ p{2.5cm}<{\centering}|m{2cm}<{\centering}|m{2cm}<{\centering}|m{2cm}<{\centering} |m{2cm}<{\centering}|m{2cm}<{\centering}}
\hline
Method & DICE (\%) & Sensitivity (\%) & Precision (\%) & VOE (\%) & RVD (\%) \\
\hline
\hline
NLC  & 60.5 $\pm$ 0.7 & 70.3 $\pm$ 2.7 & 58.8 $\pm$ 1.4 & 49.2 $\pm$ 3.4 & 30.8 $\pm$ 2.1\\
PLC  & 61.7 $\pm$ 0.7 & 76.1 $\pm$ 1.8 & 57.5 $\pm$ 0.9 & 45.9 $\pm$ 2.7 & 32.3 $\pm$ 1.3\\
\textbf{ILC}  & \textcolor{black}{\textbf{65.7 $\bm{\pm}$ 1.2}} & \textcolor{black}{\textbf{77.8 $\pm$ 2.5}} & \textcolor{black}{\textbf{60.7 $\pm$ 0.4 }}& \textcolor{black}{\textbf{44.5 $\pm$ 4.0}} & \textcolor{black}{\textbf{17.5 $\pm$ 2.3}}\\
\hline
\end{tabular}
\end{table*}
To visualize the contributions of CSAD to the prostate cancer segmentation, we compare the attention maps and the segmentation results generated by our method with and without CSAD, respectively, and the example attention maps at fifth layers are visualized in Fig.~\ref{fig_am}. 
It can be observed that with CSAD the attention maps can mainly focus on the tumor region rather than other irrelevant tissues (e.g. the top two rows of Fig. \ref{fig_am}), and the segmentation result is closer to the ground-truth compared to the ones without CSAD. This further validates that the proposed CSAD can help attention maps even at shallow layers obtain more concrete information of lesion area, and which is much helpful to subsequent attention map generation and also valuable to the final PCa segmentation task. 

\begin{table*}[h]
\newcommand{\tabincell}[2]{\begin{tabular}{@{}#1@{}}#2\end{tabular}}
\fontsize{10}{15}\selectfont
    \caption{Segmentation results of different methods on the whole MRI.}
        \label{tab2}
    \centering
  \begin{tabular}{p{3.2cm}<{\centering}|p{2cm}<{\centering}|p{2.2cm}<{\centering}|p{2cm}<{\centering}|p{2cm}<{\centering}|p{2cm}<{\centering}}
\hline
Model  & DICE (\%) & Sensitivity (\%) & Precision (\%)& VOE (\%)& RVD (\%) \\
\hline
\hline
UNet(T2W)   & 53.1 $\pm$ 1.4 & 61.4 $\pm$ 5.0 & 44.0  $\pm$ 1.9& 55.2  $\pm$ 2.3& 41.5  $\pm$ 2.7\\
UNet++(T2W)  & 54.2 $\pm$ 0.4 & 63.7 $\pm$ 3.4 & 50.1 $\pm$ 0.8 & 53.7  $\pm$ 1.9& 43.3  $\pm$ 1.4\\
UNet(ADC)   & 45.2 $\pm$ 1.7 & 60.8 $\pm$ 3.7 & 37.9 $\pm$ 1.5& 64.1  $\pm$ 3.0 & 49.1  $\pm$ 2.3\\
UNet++(ADC)   & 48.4 $\pm$ 0.6 & 61.5 $\pm$ 2.9 & 40.2 $\pm$ 0.7& 60.6  $\pm$ 2.2& 50.2  $\pm$ 1.8\\
\hline
Early Fusion & 59.3 $\pm$ 0.4 & 68.4 $\pm$ 2.3 & 57.6 $\pm$ 1.0& 50.2  $\pm$ 0.7& 28.5  $\pm$ 2.4\\
Late Fusion  & 62.2 $\pm$ 0.6 & 75.0 $\pm$ 3.7 & 60.3 $\pm$ 1.8& 47.7  $\pm$ 1.0& 24.4  $\pm$ 1.8\\
\hline
MultichannelNet\cite{Kiraly} & 56.9 $\pm$ 0.4 & 66.4 $\pm$ 1.1 & 54.3 $\pm$ 0.8& 56.2  $\pm$ 1.4& 33.3  $\pm$ 2.6\\
DeepNet\cite{AlkadiR} & 53.7 $\pm$ 1.3 & 60.8 $\pm$ 4.6 & 46.8 $\pm$ 1.6& 57.7  $\pm$ 3.7& 40.4  $\pm$ 3.9 \\
AdversarialNet\cite{Kohl} & 57.7 $\pm$ 2.4 & 69.2 $\pm$ 4.0 & 56.0 $\pm$ 3.1& 56.6  $\pm$ 4.9& 36.8 $\pm$ 2.2\\
\hline
FuseUNet\cite{LiC}  & 60.5 $\pm$ 1.4 & 69.1 $\pm$ 4.0 & 58.0 $\pm$ 1.3& 48.8  $\pm$ 0.7& 24.3  $\pm$ 1.9\\
HyperDenseNet\cite{Dolz}  & 59.4 $\pm$ 0.9 & 67.3 $\pm$ 2.6 & 58.4 $\pm$ 1.0& 53.0  $\pm$ 2.8& 24.4  $\pm$ 2.5\\
OctopusNet\cite{ChenY}  & 62.9 $\pm$ 0.8 & 75.2 $\pm$ 2.8 & 56.0 $\pm$ 2.6& 51.6  $\pm$ 1.4& 22.5  $\pm$ 2.0\\
HSCNet\cite{Jiaz} & 63.3 $\pm$ 1.1 & 74.7 $\pm$ 3.2 & 59.6 $\pm$ 1.1& 46.9  $\pm$ 2.7& 23.2  $\pm$ 3.5\\
MResUnet \cite{Ibtehaz}& 58.2 $\pm$ 0.7 & 74.3 $\pm$ 1.1 & 59.1 $\pm$ 1.4& 51.8  $\pm$ 4.2& 21.5  $\pm$ 5.4\\
\textbf{Proposed}  & \textcolor{black}{\textbf{65.7 $\bm{\pm}$ 1.2}} & \textcolor{black}{\textbf{77.8 $\bm{\pm}$ 2.5}} & \textcolor{black}{\textbf{60.7 $\bm{\pm}$ 0.4}}& \textcolor{black}{\textbf{44.5  $\pm$ 4.0}}& \textcolor{black}{\textbf{17.5  $\pm$ 2.3}} \\
\hline
\end{tabular}
\end{table*}

\begin{figure*}[h]
\setlength{\abovecaptionskip}{3pt} 
\setlength{\belowcaptionskip}{0pt}
\centerline{\includegraphics[width=\textwidth]{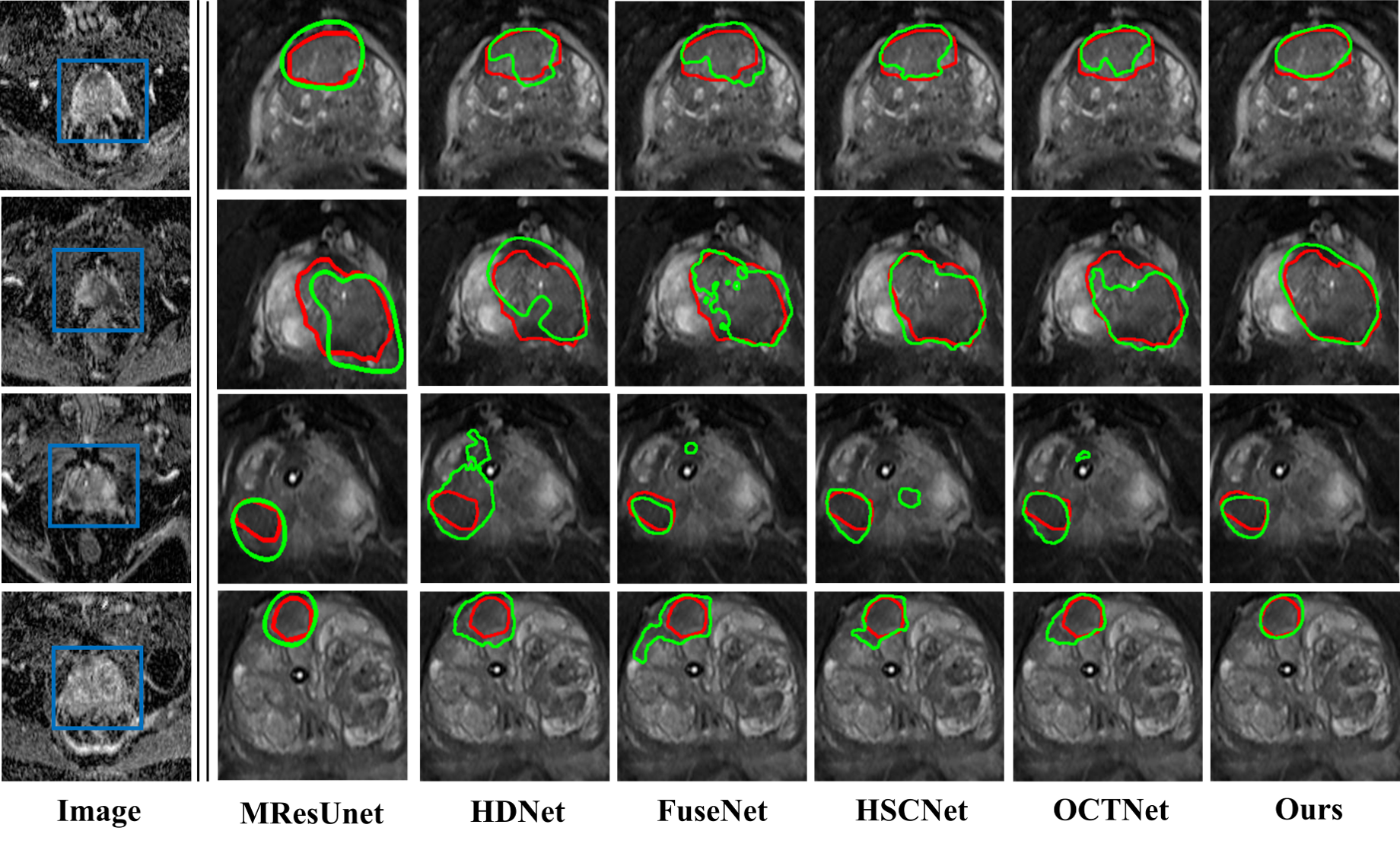}}

\caption{Segmentation results generated by different methods.
For better comparisons, we zoom in  and focus on the regions in blue bounding box of the input image. The ground-truth is marked with red, while the predicted results are marked with green.} \label{fig_seg}
\end{figure*}

\subsection{Effectiveness of Different Connection Schemes}
As mentioned in Section \ref{attention_dis},
different connection schemes of the cross attention distillations could provide various influences on the final segmentation performance.
To validate the effectiveness of different connection schemes, we perform the comparison experiments under the following three settings:
'none layer connection (\textbf{NLC})', 'position layer connection (\textbf{PLC})', 'interlaced layer connection (\textbf{ILC})'. 
Here, \textbf{NLC} is the network structure with no layer connection. \textbf{PLC} denotes the connection scheme between the same numbered layer across different modalities (i.e. $AMGB_m \rightarrow AMGB_m$), while \textbf{ILC} is the connection scheme of our proposed segmentation model, which uses the interlaced connection scheme (i.e. $AMGB_m \rightarrow AMGB_{m+1}$). Table \ref{tb_connection} provides the comparison results of different connection schemes. It is obvious that \textbf{ILC} connection scheme gains the best segmentation performance on the five comparison metrics.
There could be two reasons. Firstly, by connecting across modalities the contextual and spatial features which are significant to PCa are distilled then transferred from each modalities. Secondly, by the interlaced connection, the long dependencies of the features at each location either within a single modality or across the modalities are exploited and this is valuable to PCa. 

\subsection{Comparison with Other Methods} Furthermore, we also leverage a series of experiments to make comprehensive comparisons to the generic segmentation methods and the state-of-arts, and the quantitative results are shown in Table~\ref{tab2}. 

\textbf{Generic method.} UNet and UNet++ are two broadly used generic methods for single modality medical image segmentation. To compare their performance with ours, we trained them separately with only T2W images and only ADC images, respectively. From Table~\ref{tab2}, we can see that either UNet or UNet++ trained with T2W shows better performance than the corresponding one trained with ADC images, which reveals that T2W images can provide more useful information than ADC images do. But neither of them outperforms than our method's performance. For multi-modality UNet, we compare with FuseUNet\cite{LiC} and details can be referred to the next paragraph.
Besides, we compare the impacts of different feature fusion strategies. The early fusion network adopts the baseline architecture with the attention distillation performed within the single encoding stream, and the late fusion contains our intact structure but the SCFF module goes after the Decoder. It suggests that our proposed SCFF with intermediate layer fusion strategy is more effective than both the early fusion and the late fusion ones.

\textbf{State-of-the-arts.} For thorough comparisons, some methods particularly for the PCa segmentation or detection \cite{Kiraly} \cite{AlkadiR}  \cite{Kohl} are also re-implemented and evaluated. 
To further validate the effectiveness of our proposed CSAD network to handle the multi-modalities MRI data, other state-of-the-art multi-modalities segmentation methods not restricting to PCa are also compared: FuseUNet\cite{LiC}, HyperDenseNet\cite{Dolz}, OctopusNet\cite{ChenY}, HSCNet\cite{Jiaz}, MResUnet \cite{Ibtehaz}. Detailed comparison results are provided in Table~\ref{tab2}. And we didn’t compare with \cite{ZhangS} due to no code released and the ambiguity descriptions about the network parameters in the paper.
The result demonstrates that our proposed CSAD network yields the best segmentation results. Some segmentation results are also visualized in Fig.~\ref{fig_seg}.

\begin{table*}[h]
\newcommand{\tabincell}[2]{\begin{tabular}{@{}#1@{}}#2\end{tabular}}
\fontsize{10}{15}\selectfont
    \caption{Segmentation results on the cropped prostate region.}
        \label{tb_cropped}
    \centering
  \begin{tabular}{p{3.2cm}<{\centering}|p{2cm}<{\centering}|p{2.2cm}<{\centering}|p{2cm}<{\centering}|p{2cm}<{\centering}|p{2cm}<{\centering}}
\hline
Model  & DICE (\%) & Sensitivity (\%) & Precision (\%)& VOE (\%)& RVD (\%) \\
\hline
\hline
MResUnet \cite{Ibtehaz}& 67.5 $\pm$ 0.8 & 77.4 $\pm$ 1.2 & 73.8 $\pm$ 2.2& 35.4  $\pm$ 2.9& 14.3  $\pm$ 3.5\\
FuseUNet\cite{LiC}  & 72.0 $\pm$ 2.1 & 79.1 $\pm$ 3.1 & 75.7 $\pm$ 3.4& 31.9  $\pm$ 3.3& 6.5  $\pm$ 1.3\\
HyperDenseNet\cite{Dolz}  & 71.4 $\pm$ 0.6 & 76.3 $\pm$ 2.6 & 72.6 $\pm$ 1.7& 35.6  $\pm$ 2.4& 7.1  $\pm$ 1.9\\
OctopusNet\cite{ChenY}  & 72.8 $\pm$ 0.4 & 79.8 $\pm$ 2.4 & 79.0 $\pm$ 2.2& 31.1  $\pm$ 1.5& 6.2  $\pm$ 1.1\\
HSCNet\cite{Jiaz} & 74.2 $\pm$ 1.5 & 80.5 $\pm$ 2.7 & 78.9 $\pm$ 1.9& 33.7  $\pm$ 1.3&   $ 3.4\pm$ 1.7\\
\textbf{Proposed}  & \textcolor{black}{\textbf{76.4 $\bm{\pm}$ 1.3}} & \textcolor{black}{\textbf{81.9 $\bm{\pm}$ 1.8}} & \textcolor{black}{\textbf{80.3 $\bm{\pm}$ 1.1}}& \textcolor{black}{\textbf{30.5  $\bm{\pm}$ 2.2}}& \textcolor{black}{\textbf{1.9 $\bm{\pm}$ 0.6}} \\
\hline
\end{tabular}
\end{table*}

\subsection{Experiments on the Cropped Prostate Region}
To further validate the effectiveness of our designed CSAD model, we conduct extensive experiments on the cropped prostate region.
Compared with the segmentation task on the whole MRI, the segmentation on the cropped prostate region could be less challenging due to the smaller segmentation regions. That's the main reason that commonly low performance is gained in Table~\ref{tab2} compared to that in Table \ref{tb_cropped}.
Note that in this experiment, we also use T2W and ADC modalities, and the cropped prostate region is annotated by the radiologist manually.
The comparison results with recent state-of-the-art multi-modal segmentation methods are shown in Table \ref{tb_cropped}, it is clear that the overall performance of this task is significantly improved compared with Table \ref{tab2}.
That demonstrates that with smaller segmentation region the CSAD method could gain better performance. However, in the clinical computer aided diagnosis, the segmentation from the whole MRI could be more practical compared with the segmentation from cropped prostate region.

\section{Conclusion}
In this paper, we present a novel cross-modal self-attention distillation network for prostate cancer segmentation from a whole MRI. With the proposed CSAD network, the cross-modal reciprocal correlations are exploited and further used for good representation learning, and a SCFF module is also designed to fuse the learned features properly. The experiment results on the collected PCa MRI dataset validate the superiority of our method and also its vulnerability to other medical segmentation tasks. In the future, we will extend our CSAD network to more modalities on 2D/3D MRI at encoding/decoding stages.

\end{document}